\documentclass{article}
\usepackage{amsmath, amssymb, amsfonts}
\title{Black hole horizon around a relativistic star} 
\author{Hristu Culetu, \\Ovidius University, Dept.of Physics, \\B-dul Mamaia 124, 900527 Constanta, Romania, \\e-mail : hculetu@yahoo.com}

\begin{document}
\numberwithin{equation}{section}
\pagenumbering{arabic}
\maketitle
\newcommand{\fv}{\boldsymbol{f}}
\newcommand{\tv}{\boldsymbol{t}}
\newcommand{\gv}{\boldsymbol{g}}
\newcommand{\OV}{\boldsymbol{O}}
\newcommand{\wv}{\boldsymbol{w}}
\newcommand{\WV}{\boldsymbol{W}}
\newcommand{\NV}{\boldsymbol{N}}
\newcommand{\hv}{\boldsymbol{h}}
\newcommand{\yv}{\boldsymbol{y}}
\newcommand{\RE}{\textrm{Re}}
\newcommand{\IM}{\textrm{Im}}
\newcommand{\rot}{\textrm{rot}}
\newcommand{\dv}{\boldsymbol{d}}
\newcommand{\grad}{\textrm{grad}}
\newcommand{\Tr}{\textrm{Tr}}
\newcommand{\ua}{\uparrow}
\newcommand{\da}{\downarrow}
\newcommand{\ct}{\textrm{const}}
\newcommand{\xv}{\boldsymbol{x}}
\newcommand{\mv}{\boldsymbol{m}}
\newcommand{\rv}{\boldsymbol{r}}
\newcommand{\kv}{\boldsymbol{k}}
\newcommand{\VE}{\boldsymbol{V}}
\newcommand{\sv}{\boldsymbol{s}}
\newcommand{\RV}{\boldsymbol{R}}
\newcommand{\pv}{\boldsymbol{p}}
\newcommand{\PV}{\boldsymbol{P}}
\newcommand{\EV}{\boldsymbol{E}}
\newcommand{\DV}{\boldsymbol{D}}
\newcommand{\BV}{\boldsymbol{B}}
\newcommand{\HV}{\boldsymbol{H}}
\newcommand{\MV}{\boldsymbol{M}}
\newcommand{\be}{\begin{equation}}
\newcommand{\ee}{\end{equation}}
\newcommand{\ba}{\begin{eqnarray}}
\newcommand{\ea}{\end{eqnarray}}
\newcommand{\bq}{\begin{eqnarray*}}
\newcommand{\eq}{\end{eqnarray*}}
\newcommand{\pa}{\partial}
\newcommand{\f}{\frac}
\newcommand{\FV}{\boldsymbol{F}}
\newcommand{\ve}{\boldsymbol{v}}
\newcommand{\AV}{\boldsymbol{A}}
\newcommand{\jv}{\boldsymbol{j}}
\newcommand{\LV}{\boldsymbol{L}}
\newcommand{\SV}{\boldsymbol{S}}
\newcommand{\av}{\boldsymbol{a}}
\newcommand{\qv}{\boldsymbol{q}}
\newcommand{\QV}{\boldsymbol{Q}}
\newcommand{\ev}{\boldsymbol{e}}
\newcommand{\uv}{\boldsymbol{u}}
\newcommand{\KV}{\boldsymbol{K}}
\newcommand{\ro}{\boldsymbol{\rho}}
\newcommand{\si}{\boldsymbol{\sigma}}
\newcommand{\thv}{\boldsymbol{\theta}}
\newcommand{\bv}{\boldsymbol{b}}
\newcommand{\JV}{\boldsymbol{J}}
\newcommand{\nv}{\boldsymbol{n}}
\newcommand{\lv}{\boldsymbol{l}}
\newcommand{\om}{\boldsymbol{\omega}}
\newcommand{\Om}{\boldsymbol{\Omega}}
\newcommand{\Piv}{\boldsymbol{\Pi}}
\newcommand{\UV}{\boldsymbol{U}}
\newcommand{\iv}{\boldsymbol{i}}
\newcommand{\nuv}{\boldsymbol{\nu}}
\newcommand{\muv}{\boldsymbol{\mu}}
\newcommand{\lm}{\boldsymbol{\lambda}}
\newcommand{\Lm}{\boldsymbol{\Lambda}}
\newcommand{\opsi}{\overline{\psi}}
\renewcommand{\tan}{\textrm{tg}}
\renewcommand{\cot}{\textrm{ctg}}
\renewcommand{\sinh}{\textrm{sh}}
\renewcommand{\cosh}{\textrm{ch}}
\renewcommand{\tanh}{\textrm{th}}
\renewcommand{\coth}{\textrm{cth}}

\begin{abstract}
A star surrounded by a black hole generated by it is analyzed. For an outer static observer the black hole mass (and radius) depends on his position in the gravitational field of the star. In spite of the black hole presence, the geometry remains that of Schwarzschild. For an inner observer the black hole mass is constant and depends on the star's mass $m$ and radius $R$. The entropy for $r > R$ is one quarter of the area and is constant for $r < R$.
 
\end{abstract}

 \textbf{1. Introduction}
 
   More than a decade ago  Mannheim \cite{PM} obtained a 4-th order Einstein's equations starting with a conformally invariant Lagrangean - the Weyl tensor squared. These equations admit a static spherically symmetric vacuum solution that contains, besides the Schwarzschild term $const./r$, another term proportional to $r$ and, therefore, the metric is no longer asymptotically flat. In Mannheim's view, this linearly rising potential term shows that a local matter distribution can actually have a global effect at infinity and gravity theories become global. Using conformal invariance, he showed also that the vacuum solution has also a constant term which arises not from within a galaxy but comes from the global Hubble flow of the Universe itself.
   
  Recently Grumiller \cite{DG1} (see also \cite{RAS}) proposed a metric for gravitation at large distances, containing a new Rindler term in a spherically symmetric situation, just as Mannheim stated before in his conformally invariant model of cosmology. The Rindler constant acceleration may depend on the scale of the system under consideration and becomes important at large distances from the source. \\
  Throughout the paper we shall use the geometric units $c = G = k_{B} = \hbar = 1$.
  
  In \cite{HC1} we used the Mannheim - Grumiller geometry inside a relativistic star, where the proposed metric appears as 
\begin{equation}
 ds^{2} = -(1 - 2ar) dt^{2} + (1 - 2ar)^{-1} dr^{2} + r^{2} d \Omega^{2},
\label{1}
\end{equation}
with the constant $a$ - the intensity of the gravitational field (acceleration) anywhere its interior. Therefore, $a = m/R^{2}$, its surface value ($m$ - the star mass, $R$ - its radius) and $d\Omega^{2}$ stands for the metric on the unit 2-sphere. The spacetime (1) is a solution of Einstein's equations only if an anisotropic stress tensor of the form
 \begin{equation}
 T_{t}^{t} = - \rho = - \frac{a}{2 \pi r},~~~p_{r} = T_{r}^{r} = - \rho,~~~T^{\theta}_{\theta} = T^{\phi}_{\phi} = p_{\bot} = \frac{1}{2} p_{r}
 \label{2}
 \end{equation}
 lies on its r.h.s. In (2) $\rho$ is the energy density of the anisotropic fluid, $p_{r}$ is the radial pressure and $p_{\bot}$ are the tangential pressures. The mass till the radius $r$ is $m(r) = ar^{2} = mr^{2}/R^{2}$ \cite{HC1}. The line element (1) has an event horizon at $r = 1/2a$, located far beyond the star surface, namely in a region where the  metric (1) is not valid. \\

 \textbf{2. Outside the star}
 
 Let us now consider a static observer outside the star, at some distance $r > R$. According to the Equivalence Principle, the observer is noninertial, i.e. accelerates and, therefore, would have a horizon, exactly as the inner observer. If we take $r >> 2m$, its acceleration is given by $m/r^{2}$. We suppose that the above acceleration is rooted from a black hole (BH) of mass $M(r)$, such that
 \begin{equation}
 \frac{m}{r^{2}} = \frac{1}{4M(r)}.
\label{3}
\end{equation}
 In other words, the acceleration equals ''the surface gravity'' associated to a BH horizon. Eq. (3) yields
  \begin{equation}
  M(r) = \frac{r^{2}}{4m}.
\label{4}
\end{equation}
 We observe that $M(r) \propto r^{2}$, as we proposed recently for the BH interior \cite{HC2}. The factor $(1/4m)$ in front of $r^{2}$ plays the same role as the constant acceleration $a = m/R^{2}$ from the expression of $m(r)$ inside the star. We therefore conjecture that a static exterior observer is in fact located inside a BH with a mass given by (4) and radius $R_{g} = 2M = r^{2}/2m = r^{2}/r_{g}$, where $r_{g}$ denotes the gravitational radius associated to the star. 
 
 It is worth noting that $R_{g}$ depends on observer's position: any static observer has its own horizon. To approach the horizon the observer must remove from it. $R_{g}$ reaches its minimum value when $r = r_{min} = R$, namely when it is located on the surface. If the star becomes a BH (by collapse), one has $R = 2m$ and, therefore, $R_{g} = 2m = r_{g}$. In other words, the two horizons coincide.
 
 Let us see now whether the model is compatible with the Schwarzschild metric outside the star. We found in \cite{HC2} that the geometry inside a BH (outside the star in our situation) is given by
  \begin{equation}
   ds^{2} = -(1 - \frac{r}{2M})~ dt^{2} + (1 - \frac{r}{2M})^{-1} dr^{2} + r^{2} d \Omega^{2}~~~(r < 2M) ,
\label{5}
\end{equation}
For a fixed $r$, we put here $M = M(r)$ from (4). Hence
  \begin{equation}
   1 - \frac{r}{2M} = 1 - \frac{2m}{r}
\label{6}
\end{equation}
and the Schwarzschild metric is recovered (we have here $2m < r < 2M$). Having a Schwarzschild geometry outside the star of mass $m$ (and inside the BH of mass $M$), the stress tensor vanishes, contrary to the case from \cite{HC2}, where $M$ from Eq. (5) was constant. 

Let $M_{r}$ be the mass inside the BH, up to the radius $r > R$. Using (4) one obtains
 \begin{equation}
 M_{r} = \frac{1}{4M} r^{2} = \frac{r^{2}}{4 \frac{r^{2}}{4m}} = m ,
\label{7}
\end{equation}
namely $M_{r}$ is independent of $r$ and equals the star mass, as expected. In (7) we took advantage of the formula (6) from \cite{HC2}, i.e. $M(r) = ar^{2}$, with $a = 1/4M$, the surface gravity of the BH. Let us notice the difference between $M(r)$ and $M_{r}$: while $M(r)$ is the mass of the BH felt by an observer located at $r > R$, $M_{r}$ is the mass inside a sphere of radius $r$, calculated from $M(r)$. 

As an example, we take the case of the Sun with $m_{s} = 2.10^{33} g$ and $R_{s} = 7.10^{10} cm$, we get $R_{g} = R_{s}^{2}/r_{g} \approx 10^{16} cm$; that is, for a static observer on the Sun surface, the BH horizon finds at a distance of $10^{16} cm$ with respect to the center of the Sun (here $r_{g} \approx 3.10^{5} cm$, the Sun gravitational radius). In \cite{HC1} the above value of $R_{s}$ has been assigned to a ''Rindler'' horizon located at the distance $1/2a$.

Let us study now the thermodynamic properties of our system. As we saw in \cite{HC2}, the gravitational field inside the BH is constant and is given by its surface value $\kappa \equiv 1/4M = m/r^{2}$. Therefore, the local temperature will be 
 \begin{equation}
 T(r) = \frac{1}{8 \pi M(r)} = \frac{m}{2 \pi r^{2}} .
\label{8}
\end{equation}
The temperature is rooted from the fact that any Schwarzschild static observer is accelerated. Keeping track of all fundamental constants, we have 
 \begin{equation}
 T(r) = \frac{mc^{2}}{2 \pi k_{B}} (\frac{l_{P}}{r})^{2} ,
\label{9}
\end{equation}
where $l_{P}$ is the Planck length. From (8) we obtain $T = 1/8 \pi m = 1/8 \pi M$ when $r = 2M$ or $r = 2m$, as it should be (the star becomes a BH). Taking again the case of the Sun, one obtains for the temperature of the BH surrounding it, on the surface, $T_{S} \approx 10^{-11} K$. 

To get the expression of the entropy, we use Padmanabhan relation \cite{TP} (see also \cite{HC1}) $M_{r} = 2TS$. It yields
 \begin{equation}
 S = \frac{m}{2~T(r)} = \pi r^{2}
\label{10}
\end{equation}
which is exactly the expression obtained in \cite{HC2}, using different arguments (see also \cite{DG}).\\

\textbf{3. Inside the star}

We wish now to analyze the situation from the point of view of an observer located inside the star ($r < R$). We know from (4) that the acceleration inside is $m/R^{2} = const.$ Therefore, the BH mass will be $M(r) = R^{2}/4m = const.$ which equals the BH mass corresponding to an observer sitting on the surface. Hence, the Hawking temperature is constant inside and is given by $T_{in} = m/2 \pi R^{2}$, i.e. the temperature measured by an observer on the surface. For the entropy one obtains $S = \pi R^{2}$, a quarter of the surface area of the star. \\

\textbf{4. The entire Universe}

 Let us apply the present model to the Universe as a whole, where the metric appears as (see Eq. 5)
  \begin{equation}
   ds^{2} = -(1 - \frac{r}{2M_{U}})~ dt^{2} + (1 - \frac{r}{2M_{U}})^{-1} dr^{2} + r^{2} d \Omega^{2} ,
\label{11}
\end{equation}
 with $M_{U}$ the mass of the Universe and $a \equiv a_{U} = 1/4M_{U}$. We know from experiments that $M_{U} \approx R_{U}$, the radius of the Universe. That means an observer located far from any local mass finds himself inside a BH of mass $M_{U}$, radius $2M_{U}$ and $\kappa_{U} = 1/4M_{U}$ - the surface gravity on the horizon (or the acceleration anywhere in the interior). The horizon is located at $r = 2M_{U} = 1/2\kappa_{U} \equiv R_{U} \approx 10^{28} cm$, with respect to a chosen origin of $r$. Therefore, $\kappa_{U} \approx 10^{-8} cm/s^{2}$, which is of the order of the MOND critical acceleration $a_{0}$ \cite{MM}.
 
 We notice that the ''surface gravity'' $a_{U}$ is a scalar and, therefore, has no a certain direction. In addition, it takes the same value everywhere in empty space, far from local masses, because the cosmological horizon is not localized (it is position dependent). A similar property applies for the horizon of a BH surrounding a star: it depends on observer's position in its gravitational field. 
 
  Our goal now is to compute the active gravitational energy of the anisotropic fluid (2) for our Universe. We shall use Padmanabhan's prescription \cite{TP} (see also \cite{HC1}, Eq. 2.10) and obtain $E = - M_{U}$ (we have $a = 1/4M_{U}$ and the integration has been taken from $0$ to $2M_{U}$). We reach the same conclusion as in \cite{HC1} : the total energy (gravitational plus rest energy) of the Universe is vanishing.
 
 It is interesting to find the behaviour of a free radial photon in the metric (11). Taking $ds^{2} = 0$, one obtains
   \begin{equation}
   r(t) = R_{U} (1 - exp~(- \frac{|t|}{R_{U}})) ,~~~~~t\in(-\infty, \infty)
\label{12}
\end{equation}
In other words, the photon comes from the horizon $r = R_{U}$ at $t = - \infty$ with zero velocity, reaches $r = 0$ at $t = 0$ and goes on to $r = R_{U}$ where arrives at $t = \infty$, on the opposite side of the horizon sphere (the curve $r(t)$ is not differentiable at $t = 0$: that is related to the curvature singularity at $r = 0$). Hence, the null particle never reaches the horizon (it needs an infinite time to arrive there). That is an evidence of the BH interpretation for our Universe.\\

 \textbf{5. Conclusions}
 
 We studied the properties of the horizon a static observer outside a star would feel due to its noninertial state. The star is surrounded by a BH whose mass depends on the observer position. When it becomes a BH (because of some collapsing process), the two horizons overlap and the system will contain one BH. The Hawking temperature outside the star depends on the radial coordinate but it is constant inside.

\end{document}